\documentstyle[twoside,fleqn,espcrc2,psfig]{article}

\newcommand{\AmS}{{\protect\the\textfont2
  A\kern-.1667em\lower.5ex\hbox{M}\kern-.125emS}}
\newcommand{\lsim}{\mathrel{\mathop{\kern 0pt \rlap
  {\raise.2ex\hbox{$<$}}}
  \lower.9ex\hbox{\kern-.190em $\sim$}}}
\newcommand{\gsim}{\mathrel{\mathop{\kern 0pt \rlap
  {\raise.2ex\hbox{$>$}}}
  \lower.9ex\hbox{\kern-.190em $\sim$}}}

\title{Relic neutralinos -- update on neutralino--nucleon scalar cross-section}

\author{N. Fornengo\address{Dipartimento di Fisica Teorica,
 Universit\`a di Torino and INFN, Sezione di Torino, via
 P. Giuria 1, 10125 Torino, Italy \\
 Instituto de F\'{\i}sica 
 Corpuscular -- C.S.I.C., Departamento de F\'{\i}sica Te\`orica, 
 Universitat de Val\`encia, E-46100 Burjassot, Val\`encia, Spain \\
 {\sl fornengo@to.infn.it, fornengo@flamenco.ific.uv.es}}
 \thanks{Report on the work done in collaboration with A. Bottino,
 F. Donato and S. Scopel.}}

\begin{document}

\begin{abstract}
We discuss the effect induced on the neutralino--nucleon scalar
cross--section by the present uncertainties 
in the values of the quark masses and of the
quark scalar densities in the nucleon. 
We examine the implications of this aspect on the 
determination of the neutralino 
cosmological properties,
as derived from measurements of WIMP direct detection. We show that,  
within current theoretical uncertainties, the DAMA annual modulation 
data are compatible with a neutralino as a major dark matter 
component, to an extent which is even larger than the one previously 
derived. 
\end{abstract}

\maketitle

\begin{figure*}[t]
\hbox{
\psfig{figure=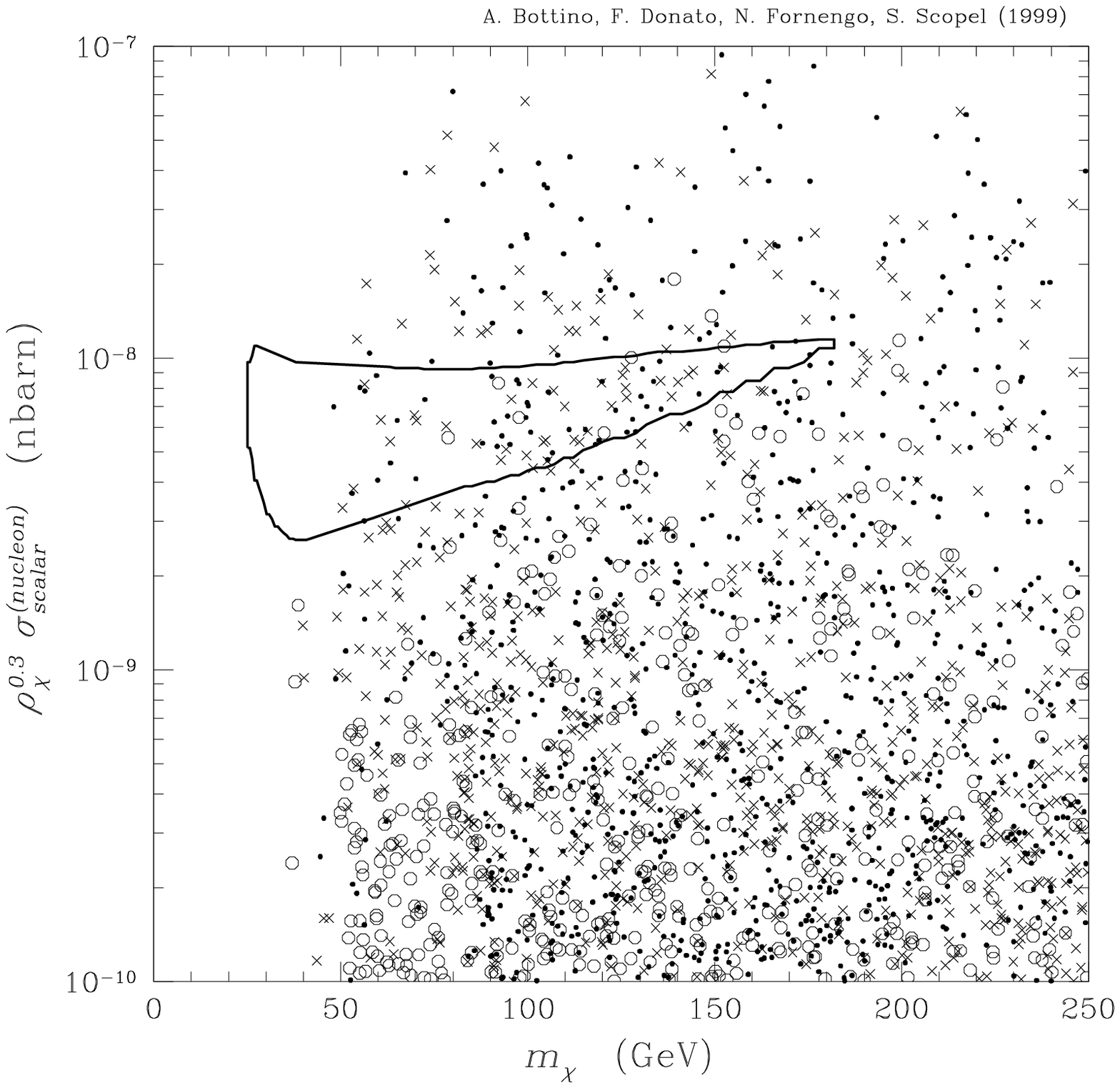,width=4.0in,bbllx=40bp,bblly=160bp,bburx=700bp,bbury=660bp,clip=}
\hspace{-70pt}
\psfig{figure=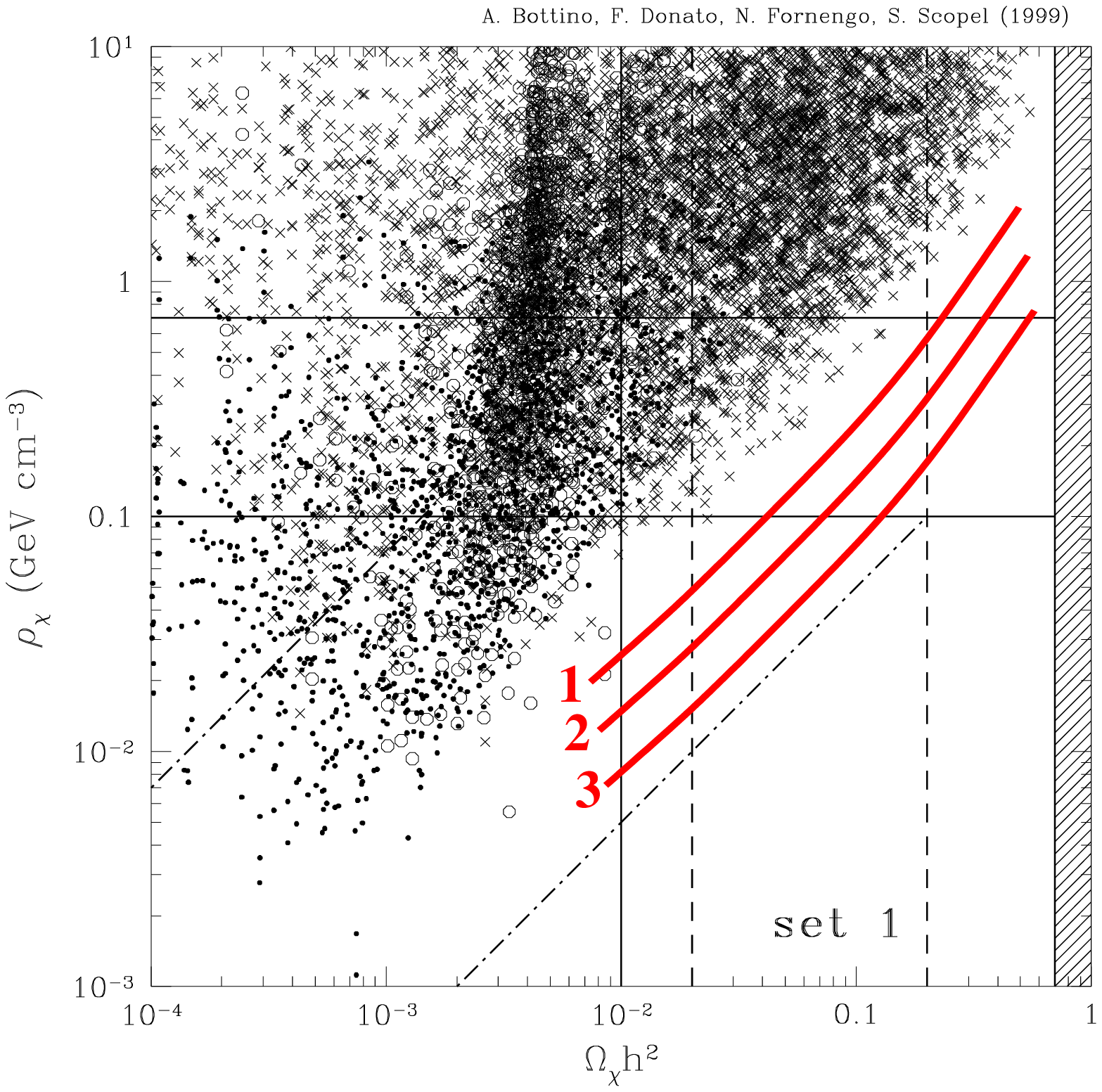,width=4.0in,bbllx=40bp,bblly=160bp,bburx=700bp,bbury=660bp,clip=}
}
\vspace{-10pt}
{Figure 1 (left). Annual--modulation region singled out by the DAMA/NaI 
experiment in the plane 
$\rho_{\chi}^{0.3} \sigma_{\mathrm scalar}^{\mathrm (nucleon)} - m_{\chi}$,
for 170 Km s$^{-1}$ $\leq v_{\mathrm loc} \leq$ 220 Km s$^{-1}$ [3].
The points represent $\sigma_{\mathrm scalar}^{\mathrm (nucleon)}$
calculated for a generic scan of the MSSM. Different symbols label
different values of the neutralino relic abundance:
$\Omega_\chi h^2 < 0.01$ (points), $0.01 \leq \Omega_\chi h^2  < 0.1$ (crosses) and
$\Omega_\chi h^2  \geq 0.1$ (circles).

Figure 2 (right). Neutralino local density $\rho_{\chi}$ derived by
requiring that $\rho_{\chi} \sigma_{\rm scalar}^{\rm (nucleon)}$ 
falls inside the experimental DAMA region of Fig. 1,
plotted against the neutralino relic abundance $\Omega_{\chi} h^2$. 
The quantity $\sigma_{\rm scalar}^{\rm (nucleon)}$ in the scatter plot
is calculated with the parameters of set 1.
The two horizontal lines delimit the physical range for the local 
density of non-baryonic dark matter. The two solid vertical lines delimit 
an interval of $\Omega_{\chi} h^2$ of cosmological interest. The 
two vertical dashed lines delimit a preferred band for cold 
dark matter. The two slant  dot--dashed lines delimit the 
band where linear rescaling procedure is usually applied. 
The shaded region is cosmologically excluded on the basis of present
limits on the age of the Universe.
Different symbols identify different neutralino
compositions: circles stand for a higgsino, crosses for a gaugino
and dots for a mixed neutralino. The solid (red) lines show to
which extent the scatter plot would enlarge if the nuclear parameters
of set 2 and of set 3, quoted in Table 1, are used.
}
\end{figure*}
\begin{table*}[hbt]
\setlength{\tabcolsep}{1.5pc}
\newlength{\digitwidth} \settowidth{\digitwidth}{\rm 0}
\catcode`?=\active \def?{\kern\digitwidth}
{Table 1. Values of the matrix elements $\langle N|\bar{q} q |N \rangle$
of the quark scalar densities in the nucleon times the quark masses $m_q$,
for different sets of values of the pion--nucleon sigma term $\sigma_{\pi N}$, 
the fractional strange--quark content of the nucleon $y$ and the quark mass
ratio $r=2m_s/(m_u+m_d)$. $q_l$ stands for light quarks, $s$ is
the strange quark and $h=c,b,t$ denotes heavy quarks. For the light
quarks, we have defined 
$m_{q_l}\langle N|\bar{q_l}q_l|N \rangle$ $\equiv$ 
$\frac{1}{2}[m_u \langle N|\bar u u|N \rangle + m_d \langle N|\bar d
d|N \rangle]$.}
\begin{tabular*}{\textwidth}{@{}l@{\extracolsep{\fill}}ccccccc}
\hline
 $\sigma_{\pi N}$ \phantom{(MeV)} & $y$  & $r$ & $m_{q_l}<\bar{q_l}q_l>$ & $m_s<\bar{s}s>$ 
& $m_h<\bar{h}h>$ &   \\  \hline
45 MeV\, & 0.33 & 29 MeV & 23 MeV & 215 MeV & 50 MeV & set 1  \\ \hline
60 MeV\, & 0.50 & 29 MeV & 30 MeV & 435 MeV & 33 MeV & set 2  \\ \hline
65 MeV\, & 0.50 & 36 MeV & 33 MeV & 585 MeV & 21 MeV & set 3  \\ \hline
\end{tabular*}
\end{table*}

Experiments of direct search for WIMPs have remarkably improved their 
sensitivity in the last years, allowing the exploration of 
sizeable regions of the physical parameter space of specific particle 
candidates for dark matter. This is the case of the neutralino \cite{bf}, 
for which some direct detection experiments are already capable of investigating  
significant features in domains of the parameter space which are also 
under current exploration at LEP2.  

Currently, the most sensitive direct search experiments are probing, 
or are starting to probe, a range of
$\rho_{\chi}^{0.3} \sigma_{\rm scalar}^{\rm (nucleon)}$ from about a few $\cdot$  
$10^{-9}$ to $1 \cdot 10^{-8}$ nbarn, where  
$\rho_{\chi}^{0.3}$ denotes the local dark matter density 
normalized to 0.3 GeV cm$^{-3}$ and $\sigma_{\rm scalar}^{\rm (nucleon)}$
is the neutralino-nucleon scalar cross section \cite{bf}.
This goal has already been achieved by the DAMA experiment 
\cite{DAMA}, which has reported the indication of an annual 
modulation effect in its counting rate, compatible with
\begin{equation} 
\label{eq:ranger}
3 \cdot 10^{-9} \; \mbox{nbarn}\lsim \rho_{\chi}^{0.3} \sigma_{\rm scalar}^{\rm (nucleon)} 
\lsim 1 \cdot 10^{-8}\; \mbox{nbarn}, 
\end{equation}
\noindent for values of the WIMP mass
which  extend over the range 
30 GeV $\lsim m_{\chi} \lsim$ 130 GeV \cite{bbbdfps}. 
The region in the  $m_{\chi}$ -- 
$\rho_{\chi}^{0.3} \sigma_{\rm scalar}^{\rm (nucleon)}$ plane, 
singled out by the DAMA experiment at 2--$\sigma$ C.L., is the one depicted 
in Fig. 1. We also notice that the uncertainties in the local 
total dark matter density: 0.1 GeV cm$^{-3} \leq \rho_l \leq $ 0.7 GeV cm$^{-3}$,
imply for $\sigma_{\rm scalar}^{\rm (nucleon)}$ the range:
\begin{equation} 
\label{eq:ranges}
1 \cdot 10^{-9} \; \mbox{nbarn}\lsim  \sigma_{\rm scalar}^{\rm (nucleon)} 
\lsim 3 \cdot 10^{-8}\; \mbox{nbarn}, 
\end{equation}
\noindent
Another experiment of WIMP direct detection which is 
now entering the upper left corner of the region in Fig. 1
is the CDMS experiment \cite{CDMS}. 

Once a given range for $\rho_{\chi}  \sigma_{\rm scalar}^{\rm (nucleon)}$ is 
singled out by an experiment, the implications for specific 
particle candidates rely on the theoretical calculation of the
$\sigma_{\rm scalar}^{\rm (nucleon)}$ cross-section. In the case
of neutralino, this quantity usually takes dominant contributions 
from interaction processes where neutralinos and quarks inside the nucleon 
interact by exchange of Higgs particles or squarks. The relevant
couplings involve the use of quark masses $m_q$ and quark 
scalar-densities inside the nucleon $\langle N|\bar{q} q |N \rangle$,
whose values can be related to some physical observables which can
be identified with the pion--nucleon sigma term $\sigma_{\pi N}$, 
the fractional strange--quark content of the nucleon $y$ and the 
strange--to--light-quark mass ratio $r=2m_s/(m_u+m_d)$ (see Ref.\cite{sets}).

Actually, the quantities $\sigma_{\pi N}$ and $y$ have recently 
been the object of various evaluations, based mainly on chiral 
perturbation theory and on QCD simulations on a lattice; however,  
the present situation, which is reviewed in Ref.\cite{sets}, 
is still far from being clear. Thus, the Higgs--quark--quark and
squark--quark--neutralino couplings are still plagued by significant 
uncertainties, and this reflects on large uncertainties on the
calculations of $\sigma_{\rm scalar}^{\rm (nucleon)}$. From
the experimentally allowed values of the relevant quantities
$\sigma_{\pi N}$, $y$ and $r$ \cite{sets}, we select three sets of values
as representative of the uncertainties which can affect the
calculation of the neutralino--nucleus cross section. These
three sets are reported in Table 1: set 1 is our reference set, 
while set 2 and set 3 are representative of choices which can
provide enhanced values for $\sigma_{\rm scalar}^{\rm (nucleon)}$.
We wish to remark that all these choices are well inside their
allowed intervals coming from the determination of the quantities
$\sigma_{\pi N}$, $y$ and $r$ \cite{sets}.

Fig. 1 shows the scatter plot of the quantity 
$\rho_{\chi}^{0.3} \sigma_{\mathrm scalar}^{\mathrm (nucleon)}$
calculated in the minimal supersymmetric standard model (MSSM), 
for a general scan of its parameter space (for details, see 
Refs. \cite{bbbdfps,sets}) and for set 1. We see that the annual
modulation region is fully compatible with the hypothesis of a
relic neutralino as a dark matter component \cite{noi1234}.

In order to discuss the astrophysical and cosmological properties 
of a relic neutralino in order for it to be compatible with
the indication coming from the DAMA experiment, in Fig. 2
we show a scatter plot of $\rho_{\chi}$ versus 
$\Omega_\chi h^2$, obtained as follows:
i) $\rho_{\chi}$ is evaluated as 
$\rho_{\chi} = 
[\rho_{\chi} \sigma_{\rm scalar}^{\rm (nucleon)}]_{R_m}/
\sigma_{\rm scalar}^{\rm (nucleon)}$, where 
$[\rho_{\chi} \sigma_{\rm scalar}^{\rm (nucleon)}]_{R_m}$ denotes the set 
of experimental values of $\rho_{\chi} \sigma_{\rm scalar}^{\rm (nucleon)}$ 
inside the DAMA annual modulation region and
$\sigma_{\rm scalar}^{\rm (nucleon)}$ is calculated in the MSSM;
ii) To each value of $\rho_{\chi}$, which then pertains to a specific 
susy configuration, one associates the corresponding 
value of $\Omega_\chi h^2$, calculated as indicated, for example,
in Ref. \cite{bf} and references therein.
With this procedure, we determine the values of $\rho_\chi$
which, for each calculated $\sigma_{\rm scalar}^{\rm (nucleon)}$
satisfy the DAMA annual modulation data.

The effect induced by the choice of different sets for the
parameters $\sigma_{\pi N}$, $y$ and $r$ is shown in Fig. 2
by the solid lines. The three lines delimit the region covered
by the scatter plots in the different cases, and show to which 
extent the region moves downward when these parameters are varied.

The most interesting feature of Fig. 2 is that it
shows that the set of susy configurations selected 
by the DAMA data has a significant overlap with the region of main 
cosmological interest:  $\Omega_\chi h^2 \gsim 0.02$ and
0.1 GeV cm$^{-3} \leq \rho_{\chi} \leq 0.7$ GeV cm$^{-3}$. 
The extent of this overlap is increasingly larger for set 2 
and set 3. By way of example, for set 3 one has that, 
at $\rho_{\chi} = 0.3$ GeV cm$^{-3}$, $\Omega_\chi h^2$ may reach the value 
0.3.  Therefore, these results reinforce our conclusions of  
Ref. \cite{noi1234,bbbdfps}, {\it i.e.} that the DAMA annual 
modulation data are compatible with a neutralino as a major component 
of dark matter, on the average in the Universe and in our Galaxy. 
 
The same figure shows that different situations may also occur.
For instance, for configurations which fall inside 
the band delimited by the slant dot--dashed lines,
the neutralino would provide only a fraction of the cold dark 
matter both at the level of local density and at the level of the 
average $\Omega$, a situation which would be possible
if the neutralino is not the unique cold dark matter particle
component. Clearly, configurations above the upper horizontal line are
incompatible with the upper limit on the local density of dark
matter in our Galaxy and must be disregarded.

\vspace{0.5cm}
{\bf Acknowledgements.}
This work was supported by the Spanish DGICYT under grant number 
PB95--1077 and by the TMR network grant ERBFMRXCT960090 of the European
Union.

\end{document}